\documentclass[%
amsmath,amssymb,
aps, reprint
]{revtex4-2}

\usepackage{hyperref} 
\usepackage{graphicx}
\usepackage{dcolumn}
\usepackage{bm}
\usepackage{graphicx}
\usepackage{epstopdf}
\usepackage{mathrsfs}
\usepackage{amssymb}
\usepackage{slashed}
\usepackage{verbatim}
\usepackage{color}
\usepackage{multirow}
\usepackage{amsmath}
\usepackage{epsfig}
\usepackage{ulem}
\usepackage{graphics}
\usepackage{indentfirst}
\usepackage{slashed}

\newcommand{\bea}{\begin{eqnarray}}
\newcommand{\eea}{\end{eqnarray}}
\newcommand{\beq}{\begin{equation}}
\newcommand{\eeq}{\end{equation}}

\def\<{\langle}
\def\>{\rangle}

\def\nn{\nonumber}

\def\cO {{\mathcal O}}

\begin{document}

\preprint{APS/123-QED}

\title{Conformality loss and short-range crossover in long-range conformal field theories} 

\author{Zhijin Li}
 \affiliation{Shing-Tung Yau Center and School of Physics, Southeast University, Nanjing, 210096, China} 

\date{\today} 

\begin{abstract}
 We study the conformality loss of theories with long-range interactions. 
     We consider the $O(2)\times O(N)$ multiscalar model with coupling $r^{-d-\delta}$ in $d=4-\epsilon$ dimension. We compute the critical exponents of the long-range fixed points (LRFPs)
         to three loops. The phase diagram of the model is dominated by two processes: the short-range crossover and merger-annihilation of LRFPs. The two processes intersect at the lower edge of the conformal window, below which the LRFPs disappear into the complex plane. We propose a novel scenario for the short-range crossover of complex LRFPs, in which the short-range crossover occurs on a vertical line in the complex plane of $\delta$. The complex marginal operator generates renormalization group flow on the transition line and significantly enriches the short-range crossover of complex LRFPs.
     
\end{abstract}
 
\maketitle


\section{Introduction}
 Conformal field theories (CFTs) with long-range interactions  \cite{Dyson:1968up, Fisher:1972zz,Sak:1973oqx, Sak:1977} play important roles in statistical mechanics \cite{Campa_2009} and quantum many-body systems \cite{Defenu:2021gzg}. The long-range interactions can be realized by defect CFTs in which the bulk free fields are coupled with operators localized on a defect \cite{Paulos:2015jfa,Caffarelli_2007}. The long-range interactions have significantly enriched the critical phenomena, and the long-range fixed points (LRFPs) show interesting differences and connections to the short-range fixed points (SRFPs). Here we focus on two aspects of LRFPs: their connection to the SRFPs and transition to the non-conformal phase, namely the loss of conformality \footnote{As will be shown later, the loss of conformality means the fixed points disappear with real couplings, but they could be reached with complex couplings. }. 

 The LRFPs with power-law decaying coupling $1/r^{d+\delta}$ are expected to transit to SRFPs for certain $\delta=\delta^*$.
 Sak suggested that for the classical $O(N)$ vector model, the long-range to short-range crossover is determined by the competition between the two types of interactions \cite{Sak:1973oqx,Sak:1977}, and the short-range crossover happens at $\delta^*=2-\eta_{\textrm{SRFP}}$, where $\eta_\textrm{SRFP}$ is the critical exponent of the SRFP. The long-range kinetic term crosses marginality condition near $\delta^*$ and the perturbative results suggest the critical exponents are continuous near the crossover. This picture has been extensively studied using different approaches \cite{Honkonen:1988fq, Honkonen:1990mr, Luijten:2002, Angelini:2014,Brezin, Giombi:2019enr,Benedetti:2020rrq,Giombi:2022gjj,Chippari:2023vme,Chippari:2023vnx, Behan:2023ile, Rong:2024vxo}.
 In \cite{Behan:2017dwr,Behan:2017emf}, it has been proposed that the crossover can be described by the SRFP perturbed by a  coupling which is weakly coupled near $\delta^*$. 

Besides the short-range crossover, evolution of the LRFPs may also end with a transition to the non-conformal phase.  A classical example is provided by the $d=1$ long-range Ising model. The model shows continuous phase transition for $\delta\in(0,1)$ \cite{Dyson:1968up} which disappears for $\delta>1$  after the short-range crossover at $\delta=1$. At the critical value $\delta=1$ the model can realize topological phase transition driven by dissociation of kink-antikink pairs \cite{Thouless:1969,Aizenman}.    
In \cite{Giachetti_2021} the authors studied the 2D $XY$ model with long-range interactions. The motivation was to dodge the Hohenberg-Mermin-Wagner theorem using long-range interactions and study their effects on the Berezinskii-Kosterlitz-Thouless topological phase transition. With long-range interactions the phase diagram of the 2D theory shows dedicated structures. 
More generally, the transition from  conformal to nonconformal phases can appear near the lower edge of the conformal window, below which the fixed points do not exist with real coupling coefficients, and the theory flows to nonconformal phases in the infrared. 

The conformal window appears in various types of theories and statistical models. They are of particular importance to determine the infrared phases of strongly coupled gauge theories. A widely interested question is through which mechanism the fixed points run out of the conformal window and disappear. It has been suggested that a pair of fixed points merge and annihilate near the lower edge of the conformal window \cite{Gies:2005as,Kaplan:2009kr}, see \cite{Gorbenko:2018dtm,Gorbenko:2018ncu} for refined discussions. Below the conformal window the fixed points disappear into the complex plane, and become a pair of complex fixed points with complex CFT data \cite{Gorbenko:2018dtm,Gorbenko:2018ncu}. The complex fixed points can be reached by tuning the complex interaction coefficients \cite{Haldar:2023ukr,Jacobsen:2024jel,Tang:2024blm}.
This mechanism is hard to verify analytically since the theories are usually strongly coupled in the regions of interest. In \cite{Benini:2019dfy}, a $d=4$ gauge theory has been proposed which is weakly coupled near the lower edge of its conformal window. The perturbative results confirmed the merger-annihilation mechanism. Likewise, the goal of this work is to establish a comprehensive picture for the conformality loss of long-range theories based on a perturbatively solvable theory,  the long-range  $O(2)\times O(N)$ multiscalar model in $d=4-\epsilon$ dimension. 
Its short-range version describes the critical behavior of the frustrated spin systems with noncollinear order \cite{Kawamura:1998}, and has been studied using perturbative and bootstrap methods with an emphasis on determining the lower edge of its conformal window in $d=3$ dimension \cite{Kawamura:1988zz,Kawamura:1990, Pelissetto:2001fi,Gracey:2002pm, Calabrese:2003ww, Nakayama:2014sba, Kompaniets_2020, Henriksson:2020fqi,Reehorst:2024vyq}. 

\begin{figure}
    \centering
    \includegraphics[width=1\linewidth]{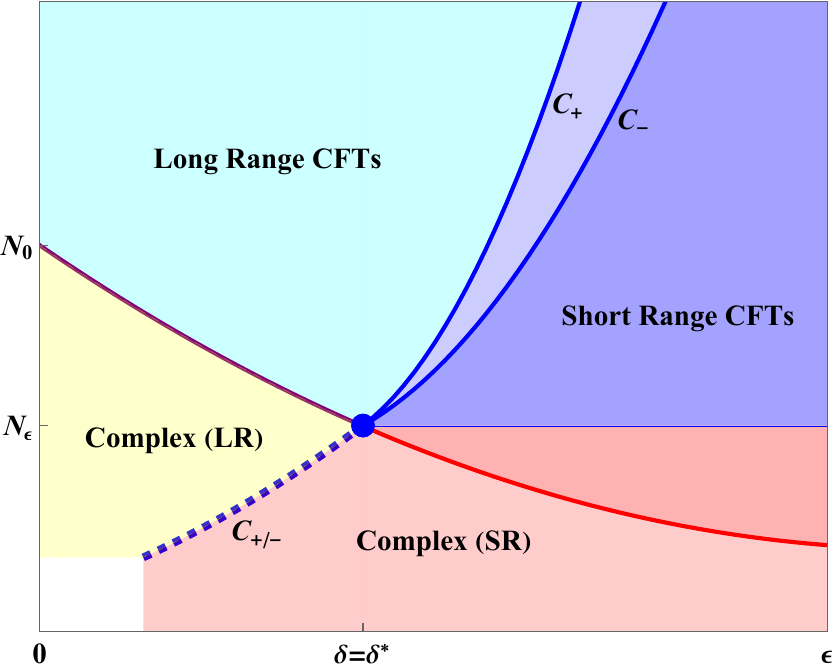}
    \caption{Phase diagram of LR $O(2)\times O(N)$ model in $d=4-\epsilon$. The purple and red lines give the lower edge of the conformal window, while the red section violates unitarity. The blue lines $C_{\pm}$ represent the SR crossover of the chiral/antichiral FPs. 
    The dashed blue line denotes  projection of SR crossover of complex LRFPs onto the real $\delta$ plane.} \label{fig:pd}
\end{figure}

Our results are summarized  in Fig. \ref{fig:pd}. Evolution of the LRFPs is dominated by two processes: the short-range crossover (blue lines) and merger-annihilation of LRFPs (purple and red lines). The two processes intersect at the lower edge of the conformal window (blue dot). 
We establish the phase diagram based on the three-loop CFT data. We find that the complex fixed points are necessary ingredients to complete this phase diagram and our analysis leads to a novel short-range crossover scenario for the complex LRFPs.
 
{\it Long-range  $O(2)\times O(N)$ multiscalar model.} We consider the long-range multi-scalar model in  $d=4-\epsilon$: 
\begin{equation}
    \mathcal{L}=\frac{1}{2}\phi_{ai}(-\partial^2)^\zeta \phi_{ai}+\lambda(\phi_{ai}\phi_{ai})^2+g\, \phi_{ai}\phi_{aj}\phi_{bi}\phi_{bj}, \label{eq:lagrangian}
\end{equation}
where $\{a,b\} /\{i,j\}$ are the $O(2)/O(N)$ indices and the summation over repeated indices is implicit.  The power of the Laplacian $\zeta\in(0,1)$ is non-integer, and it reduces to the short-range form with $\zeta\rightarrow 1$. The field $\phi$ is not renormalized due to the non-local kinetic term. Its dimension is fixed at $\Delta_\phi=(d-2\zeta)/2$. The quartic couplings are irrelevant for $\zeta< d/4$ so the theory flows to the mean field phase. For $\zeta>d/4$ the quartic couplings become relevant and generate renormalization group (RG) flow to interacting LRFPs. Near the mean field phase with $\zeta=(d+\delta)/4$, the theory can be studied perturbatively using small $\delta$ expansion.
The $\beta$ functions of this model have been computed in \cite{Benedetti:2020rrq} to three loops, which admit two fixed points related to the generalized free scalar theory and the long-range $O(2N)$ vector model. Besides, there are another two solutions for sufficiently large $N$, namely the chiral ($C_+$) and antichiral ($C_-$) fixed points. We will focus on $C_\pm$ in this work.

Our analysis on the conformality loss and short-range crossover will be based on the three-loop CFT data of the chiral/antichiral LRFPs.
We compute the scaling dimensions of the $O(2)\times O(N)$ singlet composite operators $\phi^2$ ($\Delta_{\phi^2}= d-\nu^{-1}$) and the two $\phi^4$ scalars ($\Delta_{\phi^4}= 4+\omega_{1,2}$) using the three-loop $\beta$ functions provided in \cite{Benedetti:2020rrq}. We also compute the anomalous dimensions of the long-range version of the stress tensor and $O(2)\times O(N)$ symmetry current at the LRFPs.
These perturbative results are presented in Appendices. 

{\it Short-range crossover with large $N$.} Sak's criterion \cite{Sak:1973oqx,Sak:1977} suggests the long-range to short-range crossover occurs before $\zeta=1$. The reason is that the local kinetic term $\phi\partial^2\phi$, though it does not appear in the Lagrangian (\ref{eq:lagrangian}), can be generated along the RG flow and the infrared phase is determined by the competition of the two interactions. The short-range interaction dominates with $\zeta>(2-\eta_{SR})/2$ while the long-range interaction becomes irrelevant; thus the short-range crossover is generated at $\zeta=(2-\eta_{SR})/2$, and the critical indices are expected to be continuous. However, there is spectrum decoupling and recombination at the crossover \cite{Behan:2017dwr,Behan:2017emf}. 

We apply Sak's criterion to the chiral/antichiral LRFPs. The condition $\zeta=(2-\eta_{SR})/2$ is satisfied at 
\begin{align}
    \delta^*_c&=\epsilon+\left(\frac{43}{2 n^2}-\frac{3}{2 n}\right) \epsilon ^2+\left(\frac{3}{8 n}-\frac{271}{8 n^2}\right) \epsilon ^3, \label{eq:chiralcross0} \\
\delta^*_a&=\epsilon-\frac{\epsilon ^2}{n}  +\left(\frac{2}{n^2}+\frac{1}{4 n}\right) \epsilon ^3. \label{eq:antichiralcross0}
\end{align}
for chiral/antichiral LRFPs.
We compare the critical indices of the LRFPs at the transition points $\delta_{a,c}^*$ with the short-range critical indices. 
The results confirm that the critical indices $\nu,\omega_1$ are continuous across the short-range crossover point up to three loops \footnote{The critical indices $\nu, \omega_1$ expanded in $\delta n$ are also continuous at the short-range crossover point. The critical index $\omega_2$ of the LRFPs agrees with the short-range theories at the leading order, while one has to resolve the spectrum mixing and decoupling for further consistency, the same as the long-range 3D Ising and $O(N)$ theories \cite{Benedetti:2020rrq, Rong:2024vxo}.}. 

Sak's criterion relies on the information of the SRFPs, and it will be useful to verify the short-range crossover based solely on the CFT data of LRFPs. A characteristic property of the long-range theories is the absence of conserved stress tensor, and it is expected that the anomalous dimension of the stress tensor operator vanishes at the short-range crossover \footnote{Anomalous dimensions of the stress tensor and global symmetry symmetry current in long-range Ising and $O(N)$ vector models have been computed in \cite{Behan:2017dwr,Behan:2017emf,Behan:2018hfx,Rong:2024vxo}. The results confirm that their anomalous dimensions vanish at the short-range crossover point.}. We have calculated the anomalous dimensions of the stress tensor ($\gamma_T$) and the symmetry current ($\gamma_J$) to three loops. For instance, the results for antichiral LRFP are 
\begin{align}
    \gamma_T &=\tilde{\delta} \left(\frac{1}{2}+\frac{\epsilon }{N}\right)+{\tilde{\delta}}^2 \left(\frac{19 \epsilon }{24 N}-\frac{1}{2 N}\right)- {\tilde{\delta}}^3\frac{1}{3N},\\
    \gamma_J &=\tilde{\delta} \left(\frac{1}{2}+\frac{\epsilon }{N}\right)+{\tilde{\delta}}^2 \left(\frac{5 \epsilon }{8 N}-\frac{1}{2 N}\right)- {\tilde{\delta}}^3\frac{1}{4N},
\end{align}
where $\tilde{\delta}=\delta^*_a-\delta>0$ in the long-range region. The two operators acquire positive anomalous dimensions at LRFPs. The anomalous dimensions vanish at the short-range crossover $\tilde{\delta}=0$. In the short-range region $\tilde{\delta}<0$, the stress tensor and symmetry current go below the unitary bound. While this does not necessarily mean that the theory is nonunitary, it may instead indicate that the perturbative results for (\ref{eq:lagrangian}) become invalid after the short-range crossover. The negative anomalous dimensions of the stress tensor and symmetry current are considered to be the remnants of long-range interactions.


Inspired by the work \cite{Behan:2017dwr,Behan:2017emf}, we propose a dual description of the long-range model (\ref{eq:lagrangian}) which is weakly coupled near the short-range crossover:
\begin{align}
    \mathcal{L}=&\mathcal{L}_{\textrm{SRFP}}+h\, \phi_{ai}\cdot\chi_{ai}, \label{eq:dualLag}
\end{align}
where $\mathcal{L}_{\textrm{SRFP}}$ is the short-range version of the model (\ref{eq:lagrangian}), and $\chi$ is an $O(2)\times O(N)$ bi-fundamental  generalized free scalar. By integrating out $\chi$ one can reproduce the long-range interaction in (\ref{eq:lagrangian}), given the field $\chi$ has scaling dimension $\Delta_\chi=(3d+\delta)/4$. Near the short-range crossover point $\delta^*$, the deformation $\phi\cdot\chi$ has scaling dimension $d-\left(\delta ^*-\delta \right)/4$. It crosses the marginality condition and triggers the short-range crossover at $\delta=\delta^*$. 

\vspace{1mm}
{\it Conformal window and short-range crossover.}  The chiral/antichiral LRFPs only exist with real $\lambda^*,\, g^*$ for $N\geqslant N^*$. In the merger-annihilation scenario, the conformality loss is driven by a marginal singlet scalar, which relates to the critical exponent $\omega=0$, i.e., the stability matrix  
\begin{equation}
 J(\lambda,g)\equiv 
\left(\begin{array}{cc}
  \partial_\lambda \beta_\lambda   &  \partial_g \beta_\lambda  \\
 \partial_\lambda \beta_g   &  \partial_g \beta_g
\end{array}
\right) 
\end{equation}  
acquires a vanishing eigenvalue at the fixed point:
\begin{equation}
   \left. \det[J(\lambda^*,g^*)]\right|_{N=N^*}=0.
\end{equation}
The critical value $N^*$ expanded in $\delta$ and $\epsilon=4-d$ is
\begin{align}
    N^*&=4 \left(\sqrt{6}+3\right)-\frac{1}{3} \left(7 \sqrt{6}+18\right) \delta  (2+\epsilon)+\frac{1}{900} \times \nn\\
     &\left(15 \left(47 \sqrt{6}+156\right) \zeta (3)+2198 \sqrt{6}+5772\right) \delta ^2, \label{eq:lrNc}
\end{align}
which is given by purple and red lines in Fig. \ref{fig:pd}. It intersects with the short-range crossover lines at
\begin{equation}
    \delta^*=\epsilon-\frac{1}{24}\epsilon ^2-\frac{5}{144} \epsilon ^3,
\end{equation}
for which the $N^*$ in (\ref{eq:lrNc})  exactly matches with the short-range conformal window $N_\epsilon$ in (\ref{eq:localNc}):  $$\left. N^*\right|_{\delta=\delta^*}=N_\epsilon.$$

This particular point $(\delta^*,N_\epsilon)$ is highlighted by a blue dot in Fig. \ref{fig:pd}. Both the chiral/antichiral and LRFP/SRFP collide at this critical point. There are two marginal operators in this theory that respectively generate the short-range crossover and the conformality loss processes, thus forming novel criticality. 

An interesting question is whether the long-range interactions can modify the conformal window lower than $N_\epsilon$. The result (\ref{eq:lrNc}) suggests that LRFPs still exist perturbatively for $N<N_\epsilon$ with $\delta>\delta^*$ (above red line in Fig. \ref{fig:pd}). However, with $N<N_\epsilon$ the SRFPs have already disappeared into the complex plane. This introduces a puzzle that the LRFPs can be obtained from the $\beta$ functions, while there are no corresponding SRFPs that the LRFPs can transit to!  Sak's scenario for short-range crossover simply does not apply here.

We resort to the anomalous dimensions of the stress tensor and symmetry current to verify whether the LRFPs with $N<N_\epsilon$ are physical. 
Along the lower edge of the conformal window (\ref{eq:lrNc}), the scaling dimensions of the stress tensor and symmetry current are
\begin{align}
    \Delta_T &=4-\epsilon+\tilde{\delta} \left(\frac{1}{2}+\frac{\epsilon }{24}+\frac{3 \epsilon ^2}{64}\right)+\cdots,\\
    \Delta_J &=3-\epsilon+\tilde{\delta} \left(\frac{1}{2}+\frac{\epsilon }{24}+\frac{29 \epsilon ^2}{576}\right)+\cdots,
\end{align}
where $\tilde{\delta}=\delta^*-\delta$. The scaling dimensions $\Delta_{T,J}$ go below the unitary bounds with $\delta>\delta^*$.  This suggests the LRFPs in Fig. \ref{fig:pd} with $N<N_\epsilon$ are not unitary. The negative $\gamma_{T, J}$ indicate the short-range crossover has happened before $\delta=\delta^*$.

The upshot is that the LRFPs and SRFPs share the same lower edge of conformal window at three loops. Assume this is true to all orders, it suggests the conformal window $N_\epsilon$ of the short-range model in general $4-\epsilon$ dimension can be reproduced from the $\delta$ expansion of the $d$ dimensional long-range conformal window $N^*(\delta)$, together with the short-range crossover $\gamma_T=0$. This could provide a new estimate for the physically interested short-range conformal window in $d=3$.

\vspace{1mm}
{\it Short-range crossover of complex CFTs.}
Below the conformal window, the chiral/antichiral fixed points $C_\pm$ are expected to move to the complex plane and form complex CFTs \cite{Gorbenko:2018dtm,Gorbenko:2018ncu}. 
In our case, the evolution of the complex CFTs provides an indispensable ingredient to complete the phase diagram in Fig. \ref{fig:pd}. As discussed before, the negative anomalous dimensions $\gamma_{T,J}$, considered as remnants of the long-range interaction, indicate a short-range crossover could have happened with $\delta<\delta^*$.  However, there is a contradiction for the short-range crossover of the complex LRFPs: the scaling dimension $\Delta_\phi=(d-\delta)/4$ of the LRFPs is real, while for the SRFPs, the critical indices $\eta$ below $N_\epsilon$ are complex (\ref{eq:sreta},\ref{eq:sretaN}). There is no obvious transition that can connect the two rather different critical indices. Solution to this problem leads to a new scenario for the short-range crossover of the complex LRFPs.

\begin{figure}
    \centering
    \includegraphics[width=\linewidth]{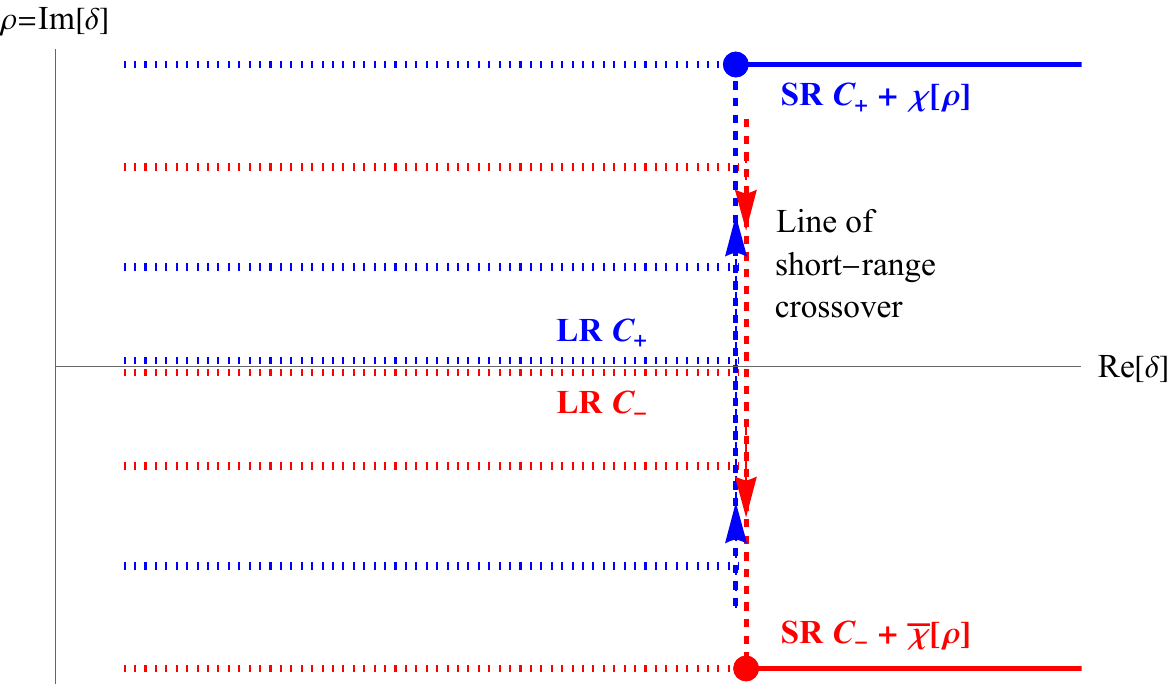}
    \caption{Phase diagram of complex CFTs with $N<N_\epsilon$. The short-range crossover occurs on the line $\textrm{Re}[\delta]=\epsilon-2\,\textrm{Re}[\eta_{SR}]$. The complex LRFPs/SRFPs locate left/right to the line.}
    \label{fig:fig2}
\end{figure}

The short-range crossover is triggered when the long-range kinetic term $\cO_K=\phi(-\partial^2)^\zeta\phi$, or $\cO_K=\phi\cdot\chi$ in (\ref{eq:dualLag}) becomes marginal $\Delta_{\cO_K}=d$, while for complex fixed points, the $\Delta_{\cO_K}$ can be a complex number for which we propose the {\it complex} marginality condition 
\begin{equation}
    \textrm{Re}[\Delta_{\cO_K}]=d. \label{complex marg}
\end{equation}
The reason is as follows. Under the RG flow, the action of this deformation $h\int d^dx \, \cO_K$ scales as 
\begin{equation}
    h\rightarrow h\mu^{-d+\Delta_{\cO_K}}=h\mu^{-d+\textrm{Re}[\Delta_{\cO_K}]} \mu^{i\,\textrm{Im}[\Delta_{\cO_K}]}.
    \end{equation}
The operator with $\textrm{Re}[\Delta_{\cO_K}]>d$ becomes irrelevant in the infrared limit $\mu\rightarrow 0$,  while its imaginary component $\textrm{Im}[\Delta_{\cO_K}]$ only contributes to an oscillating factor \footnote{Tha author would like to thank Slava Rychkov for the helpful discussion on this point.}. 
The short-range crossover of complex LRFPs occurs when the {\it complex} marginality condition is satisfied.
Note that critical indices of the chiral and antichiral SRFPs have the same real component, therefore transitions to both theories are triggered at $\textrm{Re}[\delta]=\epsilon-2\textrm{Re}[\eta_{SR}]$, denoted by vertical dashed lines in Fig. \ref{fig:fig2}. Comparing with the short-range crossover of the unitary LRFPs, the key new ingredient here is the imaginary component of $\Delta_{\cO_K}$, and the short-range crossover is extended to a transition line. Discontinuities in the critical indices during the short-range crossover of the complex LRFPs originate from the RG flow generated by the {\it complex} marginal operator $\phi\cdot\chi$ with scaling dimension $\Delta=d+i\rho$. 

We show evidence for this scenario. Near $d=4$, the short- and long-range three-loop results are valid after and before the short-range crossover, respectively. Their analytical continuations to $N<N_\epsilon$ give the critical indices of the complex fixed points. It is confirmed that at the complex transition point $\delta^*=\epsilon-2\eta_{SR}$, the short- and long-range critical indices completely agree with each other up to three loops. Moreover, the stress tensor and symmetry current are conserved at the complex transition point. The perturbative results suggest continuous short-range crossovers at complex $\delta^*$. Then we switch to the dual description (\ref{eq:dualLag}) to study the short-range crossover process with general $\delta$. The continuous transition point $\delta^*$ is reached by setting the scaling dimension of the generalized free scalar $\chi$ at $\Delta_\chi^*=(3d+\delta^*)/4$, with which the deformation $h\int d^dx\phi_{ai}\chi_{ai}$ is exactly marginal $\Delta_\phi+\Delta_\chi=d$. By varying the imaginary component of $\Delta_\chi=\Delta_\chi^*+i\rho$, above deformation becomes {\it complex} marginal. The $\beta$ function generated by such a deformation can be obtained  using the conformal perturbation theory \cite{Zamolodchikov:1987ti,Cardy:1996xt,Cappelli:1991ke,Komargodski:2016auf,Behan:2017dwr,Behan:2017emf}
\begin{equation}
    \beta_h=i\rho h+A h^3+O(h^5), \label{betah}
\end{equation}
in which the even order terms vanish due to the $O(2)\times O(N)$ symmetry and the coefficient $A$ can be extracted from the four-point function of $\cO_K$. Above $\beta$ function admits an IR fixed point at $h^2=-i\rho/A$, where the scaling dimension of the bifundamental scalar is $\Delta_\phi=(d-\delta^*)/4-i\rho$. According to our proposal for the short-range crossover, this fixed point should be the fixed point on the vertical line of short-range crossover in Fig. \ref{fig:fig2}, i.e., it contains a {\it complex} marginal operator responsible for the short-range crossover. The leading order anomalous dimension of  $\cO_K$ is purely imaginary $\gamma_\cO=\beta_h'=-2i\rho$, thus it is indeed {\it complex} marginal at the subleading order.  Our short-range crossover scenario requires such {\it complex}  marginality condition is true to all orders! This is a highly nontrivial constraint on the RG flow on the transition line, which needs to be studied further. 

In short, our proposal is that the short-range crossover for complex CFTs is controlled by the {\it complex} marginality condition (\ref{complex marg}), which corresponds to the vertical line in Fig. \ref{fig:fig2}. The LRFPs on the transition line are related to the SRFPs through RG flows generated by the {\it complex} marginal operator $\phi\cdot\chi$. This explains the discontinuities between the critical indices at the crossover. The long-range deformation $\cO_K$ becomes irrelevant in the region right to the transition line, where the SRFPs decouple from the field $\chi$. After crossing the vertical line, the LRFPs transit to SRFPs plus a family of generalized free field $\chi[\delta]$ with complex scaling dimension.
This completes the phase diagram in Fig. \ref{fig:pd}.

\vspace{1mm}
{\it Conclusion.}
We have presented a comprehensive phase diagram to show how the long-range interaction modifies the conformal window, based on the three-loop results of the long-range $O(2)\times O(N)$ multiscalar model in $d=4-\epsilon$ dimension. The phase diagram is dominated by the conformality loss and short-range crossover processes, which intersect at the lower edge of the conformal window. The long-range conformal window continues to a lower region where the unitarity has been secretly violated. Our results uncover a novel part of the phase diagram: the short-range crossover of the complex LRFPs, which generalizes the Sak's criterion for the short-range crossover to the complex plane. We proposed a new scenario that the short-range crossover occurs on a line in the complex plane labeled by the imaginary component of the {\it complex} marginal operator. We provided evidence for this scenario. The results indicate the complex CFTs have remarkable structures which are largely unexplored yet. 

The pivot of our results is the interaction between the conformality loss and short-range crossover, which is shared for general long-range theories with conformal windows, and we expect phase diagrams similar to Fig. \ref{fig:pd} can be applied for other long-range theories with conformality loss, like the $d=2$ long-range Potts model, boundary QED$_3$, etc. The complex fixed points and their short-range crossover provide a complex CFT version of the infrared duality studied in \cite{Behan:2017dwr,Behan:2017emf}.
Recently, the short-range $d=2$ complex Potts model has been studied using numerical methods \cite{Tang:2024blm,Jacobsen:2024jel}. It would be of great interest to generalize these methods to long-range cases and provide a systematical numerical study for the short-range crossover of the complex CFTs.

\vspace{3mm}
The author would like to thank Sabine Harribey, Slava Rychkov and Wei Zhu for very helpful communications and discussions. The author is grarteful to the organizers of {\it Bootstrap 2024} in {\it Universidad Complutense} in Madrid, {\it Tianfu Fields and String 2024} in Chengdu for the support during the course of this work.
This work is supported by the Startup Funding 4007022314 of the Southeast University, and the National Natural Science Foundation of China funding No. 12375061.

\bibliography{cfloss}

\clearpage

\onecolumngrid
 \begin{center}
{\bf {\large Supplementary material}}
\end{center}

\appendix
\section{Conformality loss of the short-range  $O(2)\times O(N)$ multiscalar model} \label{appendixA}
In this section, we explain the physics near the right corner $(\delta, N)=(\epsilon, N_\epsilon)$ in the phase diagram in Fig. \ref{fig:pd}. 

The short-range multi-scalar theory with an $O(2)\times O(N)$ symmetry has been studied for decades, see e.g., \cite{Kawamura:1998, Kawamura:1988zz,Kawamura:1990, Pelissetto:2001fi,Gracey:2002pm, Calabrese:2003ww, Nakayama:2014sba, Kompaniets_2020, Henriksson:2020fqi,Reehorst:2024vyq}, 
which are mainly motivated by their applications in the stacked triangular antiferromagnets. 
Actually they provide an ideal example to illustrate the vanishing of the RG fixed points. This model has been studied along this direction  even before the modern endeavors towards a general mechanism for the conformality loss \cite{Gies:2005as,Kaplan:2009kr,Gukov:2016tnp,Gorbenko:2018dtm}!
The short-range $O(2)\times O(N)$ multiscalar model provides necessary ingredients to calibrate the dynamics of its long-range version. Specifically, they relate to the physics on the vertical line at $\delta=\epsilon$ in Fig. \ref{fig:pd}. In this section, we present the three-loop results on the conformal window and critical indices of the short-range $O(2)\times O(N)$ multiscalar model, which also provides a typical example for the merger-annihilation mechanism.

The short-range multiscalar model with $O(2)\times O(N)$ symmetry is described by the Lagrangian
\begin{equation}
    \mathcal{L}=-\frac{1}{2}\phi_{ai}\partial^2 \phi_{ai}+\lambda(\phi_{ai}\phi_{ai})^2+g\, \phi_{ai}\phi_{aj}\phi_{bi}\phi_{bj}, \label{eq:srlagrangian}
\end{equation}
which corresponds to the $\zeta=1$ case of (\ref{eq:lagrangian}). The faithful global symmetry of the model is $O(2)\times O(N)/\mathbb{Z}_2$.  It has been known that its chiral and antichiral fixed points admit a conformal window with a lower edge at $N=N^*(\epsilon)$, below which the fixed points disappear into the complex plane. The conformal window of the chiral/antichiral FPs has been studied using the $4-\epsilon$ expansion up to six loops \cite{Kompaniets_2020}. The results have been resummed to evaluate the critical value $N_{\epsilon=1}$ in $d=3$ dimension, see \cite{Nakayama:2014sba,Henriksson:2020fqi,Reehorst:2024vyq} for the nonperturbative bootstrap studies of this problem. 
To clarify the long-range phase diagram in Fig. \ref{fig:pd}, it suffices to use the three-loop results \cite{Pelissetto:2001fi}, and we will focus on the theory near $d=4$, so no resummation of the perturbative results is needed.

The three-loop $\beta$ functions $\beta_\lambda, \beta_g$ of general $O(M)\times O(N)$ symmetric models are presented in \cite{Pelissetto:2001fi}. 
The equations
\begin{equation}
    \beta_\lambda(\lambda,g; N, \epsilon)=0, ~~~~ \beta_g(\lambda,g; N, \epsilon)=0, \label{eq:srbetaf}
\end{equation}
admit a pair of solutions, dubbed chiral and antichiral fixed points.
The speciality of this model is that the two $\beta$ functions (\ref{eq:srbetaf}) are endowed with a structure due to which the equations (\ref{eq:srbetaf}) can be solved with real $(\lambda,g)=(\lambda^*, g^*)$ only for $N\geqslant N_\epsilon$. The critical $N_\epsilon$ is reached at the condition 
\begin{equation}
    \det\left| \frac{\partial(\beta_\lambda,\beta_g)}{\partial(\lambda,g)}\right|_{(\lambda,g)=(\lambda^*,g^*)} =0, \label{eq:detzero}
\end{equation}
which can be solved with $(\lambda,g)=(\lambda^*, g^*)$ and $N=N^*(\epsilon)\equiv N_\epsilon$.
Physically, this suggests the critical indices $\omega=0$, and at least one of the scalar operators becomes marginal at this critical point.
The equations (\ref{eq:srbetaf}) and (\ref{eq:detzero}) together lead to a lower bound of the conformal window $N_\epsilon$ where the chiral and antichiral fixed points collide 
\begin{equation}
    N_\epsilon=4 \left(\sqrt{6}+3\right)-\frac{2}{3} \left(7 \sqrt{6}+18\right) \epsilon +\frac{1}{300} \left(274+91\sqrt{6}+ 5 \left(47 \sqrt{6}+156\right) \zeta (3)\right) \epsilon ^2 + O(\epsilon^3). \label{eq:localNc}
\end{equation}
The three-loop results on the critical indices with general $N$ are rather cumbersome. While the whole expressions are employed in our analysis of the short-range crossover, here we only show the leading terms in the large $N$ expansion for convenience.   
In the large $N$ limit, the critical indices of the chiral CFT are given by
\begin{align}
    \eta &=\left(\frac{3}{4 N}-\frac{43}{4 N^2}\right) \epsilon ^2+\left(\frac{271}{16 N^2}-\frac{3}{16 N}\right) \epsilon ^3+O(N^{-3},\epsilon^4), \label{eq:sretaN}\\
    \nu^{-1}&=2+\left(-\frac{48}{N^2}+\frac{9}{N}-1\right) \epsilon+\left(\frac{541}{4 N^2}-\frac{39}{4 N}\right) \epsilon ^2 +\left(\frac{9 (176 \zeta (3)-159)}{16 N^2}+\frac{9}{16 N}\right)\epsilon ^3+O(N^{-3},\epsilon^4), \label{eq:srnuN}\\
    \omega_1&=\left(\frac{96}{N^2}-\frac{18}{N}+1\right) \epsilon+\left(\frac{43}{2 N}-\frac{1715}{6 N^2}\right) \epsilon ^2+ \left(\frac{16151-11520 \zeta (3)}{72 N^2}-\frac{233}{72 N}\right)\epsilon ^3+O(N^{-3},\epsilon^4), \label{eq:sromega1N}\\
    \omega_2&=\epsilon+\left(\frac{821}{6 N^2}-\frac{19}{2 N}\right) \epsilon ^2+\epsilon ^3 \left(\frac{9360 \zeta (3)-14783}{72 N^2}+\frac{683}{72 N}\right)+O(N^{-3},\epsilon^4),
\end{align}
and the critical indices of the antichiral CFTs are
\begin{align}
    \eta &=\frac{1}{2 N}\epsilon ^2-\left(\frac{1}{N^2}+\frac{1}{8 N}\right) \epsilon ^3 +O(N^{-3},\epsilon^4) , \\
    \nu^{-1}&=2+\left(-\frac{36}{N^2}-\frac{6}{N}\right) \epsilon+\left(\frac{90}{N^2}+\frac{13}{2 N}\right) \epsilon ^2+\left(-\frac{121}{2 N^2}-\frac{3}{8 N}\right) \epsilon ^3+O(N^{-3},\epsilon^4), \\
    \omega_1&=\left(\frac{72}{N^2}+\frac{12}{N}-1\right) \epsilon+\left(-\frac{168}{N^2}-\frac{13}{N}\right) \epsilon ^2+\left(\frac{48 \zeta (3)+131}{N^2}+\frac{3}{4 N}\right)\epsilon ^3+O(N^{-3},\epsilon^4) , \\
    \omega_2&=\epsilon+\left(-\frac{488}{N^3}-\frac{40}{N^2}+\frac{1}{N}\right) \epsilon ^2+\left(\frac{2224}{N^3}+\frac{62}{N^2}-\frac{11}{4 N}\right) \epsilon ^3+O(N^{-3},\epsilon^4).
\end{align}

Above formulas can be used for large $N>N_0=4(\sqrt{6}+3)$ while not for the physically interested small $N<N_0$. The reason is that the critical indices solved from the perturbative $\beta$ functions contain factors proportional to $\sqrt{N-N_0}$, which becomes complex for $N<N_0$. Nevertheless, this complex factor is nonessential. It just indicates that we solved the fixed points with improper parameters. As shown in \cite{Pelissetto:2001fi,Reehorst:2024vyq}, the unphysical complex factors can be resolved by taking the parameter replacement $N=\delta n+ N_\epsilon$. The critical indices expanded in $\delta n$ and $\epsilon$ are
\begin{align}
    \eta=&\left(0.0000372659 \text{$\delta $n}^{3/2}+0.000020153 \text{$\delta $n}^2-0.000680414 \text{$\delta $n}+0.0208333\right) \epsilon ^2 \nn  \\
&+\left(0.0000516856 \text{$\delta $n}^{3/2}+0.0000658033 \text{$\delta $n}^2-0.00132681 \text{$\delta $n}+0.000436587 \sqrt{\text{$\delta $n}}+0.0173611\right) \epsilon ^3, \label{eq:sreta}\\
\nu^{-1}=&2+\left(0.00137206 \text{$\delta $n}^{3/2}-0.000201989 \text{$\delta $n}^2+0.00247449 \text{$\delta $n}-0.0903602 \sqrt{\text{$\delta $n}}-0.5\right) \epsilon  \nn\\
&+\left(0.00244777 \text{$\delta $n}^{3/2}-0.000312265 \text{$\delta $n}^2+0.000450243 \text{$\delta $n}-0.0496572 \sqrt{\text{$\delta $n}}+0.0289898\right) \epsilon ^2 \nn\\
&+\left(0.00363171 \text{$\delta $n}^{3/2}-0.000213185 \text{$\delta $n}^2-0.00439613 \text{$\delta $n}-0.0381871 \sqrt{\text{$\delta $n}}+0.0766784\right) \epsilon ^3,
\\
    \omega_1=& \left(-0.00274412 \text{$\delta $n}^{3/2}+0.000403979 \text{$\delta $n}^2-0.00494897 \text{$\delta $n}+0.18072 \sqrt{\text{$\delta $n}}\right) \epsilon \nn\\
    &+\left(-0.00541204 \text{$\delta $n}^{3/2}+0.000892423 \text{$\delta $n}^2-0.00480858 \text{$\delta $n}+0.107967 \sqrt{\text{$\delta $n}}\right) \epsilon ^2\nn\\
    &+\left(-0.00520647 \text{$\delta $n}^{3/2}+0.00170132 \text{$\delta $n}^2-0.00806641 \text{$\delta $n}+0.0603096 \sqrt{\text{$\delta $n}}\right) \epsilon ^3, \label{eq:sromega1}\\
    \omega_2=&  \epsilon+ \left(0.00252367 \text{$\delta $n}^{3/2}-0.000497326 \text{$\delta $n}^2+0.00806582 \text{$\delta $n}-0.0419124 \sqrt{\text{$\delta $n}}-0.159041\right) \epsilon ^2 \nn\\
    &+\left(-0.000637935 \text{$\delta $n}^{3/2}-0.000243656 \text{$\delta $n}^2-0.000651305 \text{$\delta $n}+0.0259794 \sqrt{\text{$\delta $n}}+0.077255\right) \epsilon ^3 
    , \label{eq:sromega2}
\end{align}
for the chiral fixed points and
\begin{align}
    \eta=&\left(-0.0000372659 \text{$\delta $n}^{3/2}+0.000020153 \text{$\delta $n}^2-0.000680414 \text{$\delta $n}+0.0208333\right) \epsilon ^2 \nn\\
    &+\left(-0.0000516856 \text{$\delta $n}^{3/2}+0.0000658033 \text{$\delta $n}^2-0.00132681 \text{$\delta $n}-0.000436587 \sqrt{\text{$\delta $n}}+0.0173611\right) \epsilon ^3,  \label{etaDN}\\
    \nu^{-1}=&2+ \left(-0.00137206 \text{$\delta $n}^{3/2}-0.000201989 \text{$\delta $n}^2+0.00247449 \text{$\delta $n}+0.0903602 \sqrt{\text{$\delta $n}}-0.5\right) \epsilon \nn\\
    &+\left(-0.00244777 \text{$\delta $n}^{3/2}-0.000312265 \text{$\delta $n}^2+0.000450243 \text{$\delta $n}+0.0496572 \sqrt{\text{$\delta $n}}+0.0289898\right) \epsilon ^2\nn\\
    &+\left(-0.00363171 \text{$\delta $n}^{3/2}-0.000213185 \text{$\delta $n}^2-0.00439613 \text{$\delta $n}+0.0381871 \sqrt{\text{$\delta $n}}+0.0766784\right) \epsilon ^3,
\\
    \omega_1=&\left(0.00274412 \text{$\delta $n}^{3/2}+0.000403979 \text{$\delta $n}^2-0.00494897 \text{$\delta $n}-0.18072 \sqrt{\text{$\delta $n}}\right) \epsilon\nn\\
    &+\left(0.00541204 \text{$\delta $n}^{3/2}+0.000892423 \text{$\delta $n}^2-0.00480858 \text{$\delta $n}-0.107967 \sqrt{\text{$\delta $n}}\right) \epsilon ^2\nn\\
    &+\left(0.00520647 \text{$\delta $n}^{3/2}+0.00170132 \text{$\delta $n}^2-0.00806641 \text{$\delta $n}-0.0603096 \sqrt{\text{$\delta $n}}\right) \epsilon ^3, \\
\omega_2=&\epsilon+\left(-0.00252367 \text{$\delta $n}^{3/2}-0.000497326 \text{$\delta $n}^2+0.00806582 \text{$\delta $n}+0.0419124 \sqrt{\text{$\delta $n}}-0.159041\right) \epsilon ^2 \nn\\
    &+\left(0.000637935 \text{$\delta $n}^{3/2}-0.000243656 \text{$\delta $n}^2-0.000651305 \text{$\delta $n}-0.0259794 \sqrt{\text{$\delta $n}}+0.077255\right) \epsilon ^3, \label{w2DN}
\end{align}
for the antichiral fixed points. 
Here we have implicitly assumed the higher order corrections in the $\epsilon$ and $\delta n$ expansions.

Note that the perturbative results (\ref{etaDN}-\ref{w2DN}) are actually expanded in $\sqrt{\delta n}=\sqrt{N-N_\epsilon}$, and the only difference between the two sets of critical indices is that they have opposite signs for the odd powers of $\sqrt{\delta n}$. There are three regions in the parameter space $N$ corresponding to different stages of the merger-annihilation process:
\begin{itemize}
    \item $N>N_\epsilon$, $\delta n>0$:  the critical indices (\ref{etaDN}-\ref{w2DN}) are real and the theories are unitary.  The leading term of $\omega_1$ is positive in chiral CFT $\sim 0.18072\sqrt{\delta n}\epsilon$ while is negative in antichiral CFT $\sim -0.18072\sqrt{\delta n}\epsilon$. The two theories are stable/unstable in this region.
    \item $N=N_\epsilon$, $\delta n=0$: there are no differences between the chiral and antichiral fixed points. The two theories merge with each other. The critical index $\omega_1=0$, suggesting a marginal singlet scalar at this particular point. This is the smoking gun of the merger-annihilation scenario.
    \item $N<N_\epsilon$, $\delta n<0$:  the critical indices (\ref{etaDN}-\ref{w2DN}) become complex. The chiral/antichiral fixed points have exactly the same real component while opposite imaginary component, i.e., the two sets of critical indices are complex conjugate of each other. 
\end{itemize}
Above results provide a comprehensive picture for the merger-annihilation mechanism for the conformality loss. The $\beta$ functions near $N_\epsilon$ can be described by the saddle-node bifurcation theory \cite{Gukov:2016tnp}.

Nevertheless, the ``annihilation" at $N_\epsilon$ is not the end of the fixed points. These fixed points are expected to become a pair of complex CFTs below the conformal window $N<N_\epsilon$ \cite{Gorbenko:2018dtm,Gorbenko:2018ncu}. Strong evidence of the  complex CFTs has been observed in recent numerical simulations \cite{Haldar:2023ukr,Tang:2024blm,Jacobsen:2024jel}. 

\section{Perturbative results for the long-range  $O(2)\times O(N)$ multiscalar model} \label{appendixB}
In this section, we show the details of the perturbative results for the long-range fixed points, which provide substantial ingredients to establish the phase diagram in Fig. \ref{fig:pd}.

We compute the critical indices of the $O(2)\times O(N)$ long-range chiral/antichiral FPs based on the three-loop $\beta$ functions obtained in \cite{Benedetti:2020rrq}, in which the BPHZ (Bogoliubov-Parasiuk-Hepp-Zimmermann) subtraction scheme has been employed. The $\beta$ functions are expanded in $\delta=4\zeta-d$ in general dimension $d<4$. The long-range kinetic term in (\ref{eq:lagrangian}) generates a propagator with a non-integer power, which makes the Feynman diagrams of long-range theories much harder than the SR cases. To overcome this problem, the authors in \cite{Benedetti:2020rrq} employed the Mellin-Barnes representation of Feynman diagrams in the Schwinger parametrization. The method developed in \cite{Benedetti:2020rrq} can be applied to scalar theories with general quartic interactions, and by imposing the $O(2)\times O(N)$ symmetry it gives results for the  chiral model (\ref{eq:lagrangian}). 

Another important data for the long-range theories is the scaling dimensions of the stress tensor and global symmetry current. A symbolic property of the long-range theories is that the stress tensor acquires anomalous dimensions and is not conserved. The anomalous dimensions of the stress tensor and symmetry current have been computed in \cite{Behan:2017dwr,Behan:2017emf,Behan:2018hfx,Rong:2024vxo} for the long-range $O(N)$ vector models, which confirm that at the crossover point $\delta=\delta^*$, the anomalous dimensions vanish and the conservation conditions are restored, consistent with the Sak's scenario for the short-range crossover \cite{Sak:1973oqx}.  In \cite{Behan:2017dwr,Behan:2017emf,Behan:2018hfx,Behan:2023ile} the anomalous dimension of the stress tensor is obtained with conformal perturbation and bootstrap methods, while in \cite{Rong:2024vxo} the author adopted the Feynman diagram method developed in \cite{Benedetti:2020rrq}. We follow the method in \cite{Benedetti:2020rrq} to compute the anomalous dimensions of the stress tensor and symmetry current, similar to \cite{Rong:2024vxo} but with a different global symmetry in the  scalar quartic interactions. Here we present the final results that will be used in this work, and refer to \cite{Benedetti:2020rrq}
for further explanations.

\vspace{3mm}
{\noindent\bf Conformal window:} The $\beta$ functions of the couplings $\lambda$ and $g$ in the long-range theory (\ref{eq:lagrangian}) have chiral and antichiral LRFPs, given $N$ above a $\delta$-dependent critical value $N_\delta$:
\begin{align}
    N_\delta= & 4 \left(\sqrt{6}+3\right)+\frac{2\delta}{3} \left(7 \sqrt{6}+18\right)   \left(-\psi ^{(0)}\left(\frac{d}{2}\right)+2 \psi ^{(0)}\left(\frac{d}{4}\right)+\gamma \right) +\frac{\delta ^2}{3600}\left\{10802 \gamma ^2 \sqrt{6}+27768 \gamma ^2\right. \nn\\
& +\sqrt{6} \left(2 \left(2314 \sqrt{6}+5401\right) \left(2 \psi ^{(0)}\left(\frac{d}{4}\right)-\psi ^{(0)}\left(\frac{d}{2}\right)\right) \left(2 \left(\psi ^{(0)}\left(\frac{d}{4}\right)+\gamma \right)-\psi ^{(0)}\left(\frac{d}{2}\right)\right)\right) \nn\\
& \left.+\frac{5 \left(47 \sqrt{6}+156\right) \Gamma \left(\frac{d}{4}+1\right)^3 \Gamma \left(-\frac{d}{4}\right) \left(\pi ^2-6 \psi ^{(1)}\left(\frac{d}{4}\right)\right)}{\Gamma \left(\frac{d}{2}\right)}\left.-5 \left(26 \sqrt{6}+67\right) \left(\pi ^2-6 \psi ^{(1)}\left(\frac{d}{2}\right)\right)\right)\right\}.
\end{align}
Note the long-range interaction parameter $\delta\in(0,\epsilon)$, and above critical value $N_\delta$ exists perturbatively for any $\delta$ in this range. Its relation to the short-range critical value $N_\epsilon$ in (\ref{eq:localNc}) is studied carefully in the main body of this paper.

\vspace{3mm}
{\noindent\bf Critical indices:}
The critical index $\eta$ is non-renormalized due to the non-local kinetic term in (\ref{eq:lagrangian}), so we have 
\begin{equation}
    \eta=2(1-\zeta)=(4-d-\delta)/2.
\end{equation}
In contrast, the composite operators, including the stress tensor and symmetry current do acquire anomalous dimensions. The whole perturbative expansions for the critical indices are rather cumbersome. Below we present the results by further taking the large $N$ expansion or the truncated expansions of $N-N_\delta$. Nevertheless, the conclusions made in the main body are established on the exact three-loop perturbative results. 

In the large $N$ limit, the $\epsilon$ expansions of the critical indices of the chiral LRFPs  are given by
\begin{align}
    \nu^{-1}=&\frac{d}{2}-\frac{  \left(N^2-18 N+96\right)}{2 N^2}\delta+\frac{(21 N-319) \left(-\psi ^{(0)}\left(\frac{d}{2}\right)+2 \psi ^{(0)}\left(\frac{d}{4}\right)+\gamma \right)}{2 N^2}\delta ^2 +\frac{1}{16 N^2} \delta ^3\left\{ 2 \gamma ^2 (57 N-2059)   \right.\nn\\
   &\left. +\pi ^2 (97-3 N) +2 (57 N-2059) \left(2 \psi ^{(0)}\left(\frac{d}{4}\right)-\psi ^{(0)}\left(\frac{d}{2}\right)\right) \left(2 \left(\psi ^{(0)}\left(\frac{d}{4}\right)+\gamma \right)-\psi ^{(0)}\left(\frac{d}{2}\right)\right)\right. \nn\\
   &\left. +18 N \psi ^{(1)}\left(\frac{d}{2}\right)-582 \psi ^{(1)}\left(\frac{d}{2}\right)-\frac{132 \pi  \csc \left(\frac{\pi  d}{4}\right) \Gamma \left(\frac{d}{4}+1\right)^2 \left(\pi ^2-6 \psi ^{(1)}\left(\frac{d}{4}\right)\right)}{\Gamma \left(\frac{d}{2}\right)} \right\}, \\
   \omega_1=&\frac{\delta  ((N-18) N+96)}{N^2}-\frac{\delta ^2 (69 N-1003) \left(-\psi ^{(0)}\left(\frac{d}{2}\right)+2 \psi ^{(0)}\left(\frac{d}{4}\right)+\gamma \right)}{3 N^2}+\frac{\delta ^3 }{72 N^2}\left\{ \pi ^2 (33 N-931)  \right. \nn\\
   &\left.    -2 (643 N-20833) \left(2 \psi ^{(0)}\left(\frac{d}{4}\right)-\psi ^{(0)}\left(\frac{d}{2}\right)\right) \left(2 \left(\psi ^{(0)}\left(\frac{d}{4}\right)+\gamma \right)-\psi ^{(0)}\left(\frac{d}{2}\right)\right)-198 N \psi ^{(1)}\left(\frac{d}{2}\right) \right. ~~~~~~ \nn\\
   &\left.+2 \gamma ^2 (20833-643 N) +5586 \psi ^{(1)}\left(\frac{d}{2}\right)    +\frac{960 \pi  \csc \left(\frac{\pi  d}{4}\right) \Gamma \left(\frac{d}{4}+1\right)^2 \left(\pi ^2-6 \psi ^{(1)}\left(\frac{d}{4}\right)\right)}{\Gamma \left(\frac{d}{2}\right)}\right\}, \\
   \omega_2=& \frac{\delta ^2 (33 N-475) \left(-\psi ^{(0)}\left(\frac{d}{2}\right)+2 \psi ^{(0)}\left(\frac{d}{4}\right)+\gamma \right)}{3 N^2}+\delta +\frac{1}{72 N^2} \delta ^3 \left\{ 2 \gamma ^2 (733 N-17107)+\pi ^2 (841-15 N)   \right. \nn\\
   & + 2 (733 N-17107) \left(2 \psi ^{(0)}\left(\frac{d}{4}\right)-\psi ^{(0)}\left(\frac{d}{2}\right)\right) \left(2 \left(\psi ^{(0)}\left(\frac{d}{4}\right)+\gamma \right)-\psi ^{(0)}\left(\frac{d}{2}\right)\right)-5046 \psi ^{(1)}\left(\frac{d}{2}\right) \nn \\
   &\left. +90 N \psi ^{(1)}\left(\frac{d}{2}\right)+\frac{72 (3 N-43) \Gamma \left(-\frac{d}{4}\right) \Gamma \left(\frac{d}{2}\right)^2}{\Gamma \left(\frac{3 d}{4}\right)} +\frac{780 \Gamma \left(\frac{d}{4}+1\right)^3 \Gamma \left(-\frac{d}{4}\right) \left(\pi ^2-6 \psi ^{(1)}\left(\frac{d}{4}\right)\right)}{\Gamma \left(\frac{d}{2}\right)}\right\},
\end{align}
and the critical indices of the antichiral LRFPs are
\begin{align}
    \nu^{-1}=&\frac{d}{2}+\delta  \left(\frac{1}{2}-\frac{6 (N+6)}{N^2}\right)-\frac{7 \delta ^2 (N+12) \left(-\psi ^{(0)}\left(\frac{d}{2}\right)+2 \psi ^{(0)}\left(\frac{d}{4}\right)+\gamma \right)}{N^2}-\frac{\delta ^3}{8 N^2}\left\{  \right.\gamma ^2 (38 N+848) \nn \\
    & +2 (19 N+424) \left(2 \psi ^{(0)}\left(\frac{d}{4}\right)-\psi ^{(0)}\left(\frac{d}{2}\right)\right) \left(2 \left(\psi ^{(0)}\left(\frac{d}{4}\right)+\gamma \right)-\psi ^{(0)}\left(\frac{d}{2}\right)\right) \nn \\
    &\left. -\pi ^2 (N+24) +6 N \psi ^{(1)}\left(\frac{d}{2}\right)+144 \psi ^{(1)}\left(\frac{d}{2}\right)\right\},    \\   
    \omega_1=&\delta  \left(\frac{12 (N+6)}{N^2}-1\right)+\frac{2 \delta ^2 (7 N+78) \left(-\psi ^{(0)}\left(\frac{d}{2}\right)+2 \psi ^{(0)}\left(\frac{d}{4}\right)+\gamma \right)}{N^2}+\frac{\delta ^3}{4 N^2} \left\{ \gamma ^2 (38 N+804) \right. \nn\\
    &\left. +2 (19 N+402) \left(2 \psi ^{(0)}\left(\frac{d}{4}\right)-\psi ^{(0)}\left(\frac{d}{2}\right)\right) \left(2 \left(\psi ^{(0)}\left(\frac{d}{4}\right)+\gamma \right)-\psi ^{(0)}\left(\frac{d}{2}\right)\right)+6 N \psi ^{(1)}\left(\frac{d}{2}\right)\right.\nn\\
    &\left. -\pi ^2 (N+14)+84 \psi ^{(1)}\left(\frac{d}{2}\right)-\frac{16 \pi  \csc \left(\frac{\pi  d}{4}\right) \Gamma \left(\frac{d}{4}+1\right)^2 \left(\pi ^2-6 \psi ^{(1)}\left(\frac{d}{4}\right)\right)}{\Gamma \left(\frac{d}{2}\right)} \right\}, \\
    \omega_2=& \delta+ \frac{40 \delta ^2 \left(-\psi ^{(0)}\left(\frac{d}{2}\right)+2 \psi ^{(0)}\left(\frac{d}{4}\right)+\gamma \right)}{N^2}+ \frac{2 \delta ^3}{N^2} \left\{ 48 \gamma ^2-\pi ^2++6 \psi ^{(1)}\left(\frac{d}{2}\right) +\frac{N \Gamma \left(-\frac{d}{4}\right) \Gamma \left(\frac{d}{2}\right)^2}{\Gamma \left(\frac{3 d}{4}\right)}\right. \nn\\
    &\left. +48 \left(2 \psi ^{(0)}\left(\frac{d}{4}\right)-\psi ^{(0)}\left(\frac{d}{2}\right)\right) \left(2 \left(\psi ^{(0)}\left(\frac{d}{4}\right)+\gamma \right)-\psi ^{(0)}\left(\frac{d}{2}\right)\right) \right\}.
\end{align}
The short-range crossover is triggered at 
\begin{equation}
    \delta^*=\epsilon+\left(\frac{43}{2 N^2}-\frac{3}{2 N}\right) \epsilon ^2+\left(\frac{3}{8 N}-\frac{271}{8 N^2}\right) \epsilon ^3+\cdots \label{eq:chiralcross}
\end{equation}
for chiral LRFP and at
\begin{equation}
\delta^*=\epsilon-\frac{\epsilon ^2}{N}  +\left(\frac{2}{N^2}+\frac{1}{4 N}\right) \epsilon ^3  +\cdots \label{eq:antichiralcross}
\end{equation}
for antichiral LRFP.

For the more physically interesting case with $N<N_0=4(\sqrt{6}+3)$, due to the same reason for the short-range case, the critical indices need to be expanded with respect to $\delta n=N-N_\delta$. We show the critical indices of the  chiral LRFP
\begin{align}
    \nu^{-1}=& 2-0.5\epsilon+\delta  \left(0.00137206 \text{$\delta $n}^{3/2}-0.000201989 \text{$\delta $n}^2+0.00247449 \text{$\delta $n}-0.0903602 \sqrt{\text{$\delta $n}}\right) \nn\\
    &+\delta ^2 \left(0.0026279 \text{$\delta $n}^{3/2}-0.000330783 \text{$\delta $n}^2+0.000553347 \text{$\delta $n}-0.0534222 \sqrt{\text{$\delta $n}}+0.0289898\right) \nn\\
    &+\delta ^3 \left(0.00287208 \text{$\delta $n}^{3/2}-0.0000777628 \text{$\delta $n}^2-0.00485637 \text{$\delta $n}-0.0176248 \sqrt{\text{$\delta $n}}+0.0645993\right) \nn\\
    &+\delta^2\epsilon\left(0.00131395 \text{$\delta $n}^{3/2}-0.000165392 \text{$\delta $n}^2+0.000276673 \text{$\delta $n}-0.0267111 \sqrt{\text{$\delta $n}}+0.0144949\right)  , \label{lrnu}\\
    \omega_1=& \delta  \left(-0.00274412 \text{$\delta $n}^{3/2}+0.000403979 \text{$\delta $n}^2-0.00494897 \text{$\delta $n}+0.18072 \sqrt{\text{$\delta $n}}\right) \nn\\
    &+\delta ^2 \left(-0.00577231 \text{$\delta $n}^{3/2}+0.00092946 \text{$\delta $n}^2-0.00501479 \text{$\delta $n}+0.115497 \sqrt{\text{$\delta $n}}\right) \nn\\
    &+ \delta^2\epsilon\left(-0.00288615 \text{$\delta $n}^{3/2}+0.00046473 \text{$\delta $n}^2-0.0025074 \text{$\delta $n}+0.0577485 \sqrt{\text{$\delta $n}}\right), \label{lromega1}\\
    \omega_2=& \delta+\delta ^2 \left(0.0419124 \sqrt{\text{$\delta $n}}-0.200707\right)+\cdots.
\end{align}
The critical indices for the antichiral LRFP can be obtained by flipping the signs of the terms with odd power of $\sqrt{\delta n}$. The short-range crossover of the chiral LRFP occurs at 
\begin{align}
\delta=\delta^*=&\epsilon+\left(-0.0000745317 \text{$\delta $n}^{3/2}-0.000040306 \text{$\delta $n}^2+0.00136083 \text{$\delta $n}-0.0416667\right) 
    \epsilon ^2 \nn\\
    &+\left(-0.000103371 \text{$\delta $n}^{3/2}-0.000131607 \text{$\delta $n}^2+0.00265361 \text{$\delta $n}-0.000873175 \sqrt{\text{$\delta $n}}-0.0347222\right) \epsilon ^3. \label{eq:srcrossoverdeltaN}
\end{align}
At the crossover point $\delta^*$, the long-range critical indices $\nu^{-1}, \omega_1$ continuously transit to the short-range version (\ref{eq:sreta}-\ref{eq:sromega1}), while the $\omega_2$ differs at the subleading order, which is not surprising as the long-range sector contains more spectrum than the short-range case \cite{Behan:2017dwr,Behan:2017emf}.

\vspace{3mm}
{\noindent\bf Anomalous dimensions of stress tensor and symmetry current:} The anomalous dimension of the stress tensor and symmetry current play two roles in this work: {\bf (a)} verify the locality, or the short-range crossover of the theory and {\bf (b)} check if the theory is unitary. Their anomalous dimensions, together with the critical indices provide strong evidence to establish the phase diagram in Fig. \ref{fig:pd}. 

We first show the scaling dimensions of the stress tensor $\Delta_T$ and the $O(2)\times O(N)$ symmetry current $\Delta_J$ with large $N$.
It is convenient to shift the expansion variable $\delta\rightarrow \tilde{\delta}=\delta^*-\delta$, where $\delta^*$ is the short-range crossover point given by (\ref{eq:chiralcross}) for chiral fixed point and (\ref{eq:antichiralcross}) for antichiral fixed point, or (\ref{eq:srcrossoverdeltaN}) if expanded near $N_\delta$.
The results are
\begin{align}
    \Delta_T &=4-\epsilon+\tilde{\delta} \left(\frac{1}{2}+\frac{(3 N-43) \epsilon }{2 N^2}-\frac{7 (3 N-205) \epsilon ^2}{24 N^2}\right)+{\tilde{\delta}}^2 \left(\frac{(57 N-3193) \epsilon }{48 N^2}+\frac{43-3 N}{4 N^2}\right)+{\tilde{\delta}}^3\frac{ (142-3 N)}{6 N^2} ,\\
    \Delta_J &=3-\epsilon+\tilde{\delta} \left(\frac{1}{2}+\frac{(3 N-43) \epsilon }{2 N^2}+\frac{(232-3 N) \epsilon ^2}{4 N^2}\right)+{\tilde{\delta}}^2 \left(\frac{(15 N-1007) \epsilon }{16 N^2}+\frac{43-3 N}{4 N^2}\right)+{\tilde{\delta}}^3\frac{(175-3 N)}{8 N^2} ,
\end{align}
for the chiral fixed point and
\begin{align}
    \Delta_T &=4-\epsilon+\tilde{\delta} \left(-\frac{(7 N+12) \epsilon ^2}{12 N^2}+\frac{\epsilon }{N}+\frac{1}{2}\right)+{\tilde{\delta}}^2 \left(\frac{19 \epsilon }{24 N}-\frac{1}{2 N}\right)- {\tilde{\delta}}^3\frac{1}{3N},\\
    \Delta_J &=3-\epsilon+\tilde{\delta} \left(-\frac{(N+2) \epsilon ^2}{2 N^2}+\frac{\epsilon }{N}+\frac{1}{2}\right)+{\tilde{\delta}}^2 \left(\frac{5 \epsilon }{8 N}-\frac{1}{2 N}\right)- {\tilde{\delta}}^3\frac{1}{4N}.
\end{align}
The above results show two interesting facts: {\bf (a)} at the short-range crossover point $\delta=\delta^*$, the anomalous dimensions of the stress tensor and symmetry current vanish; {\bf (b)} in the long-range region $\delta<\delta^*$, scaling dimensions of the stress tensor and symmetry current are above the unitary bound, while they are lower than the unitary bounds after the short-range crossover ($\delta>\delta^*$). This does not necessarily mean that the theory becomes nonunitary; instead, it may indicates that the long-range feature of the theory has been modified. 

Then we consider the stress tensor and symmetry current along the line where the chiral/antichiral LRFPs collide (the purple line and red line in Fig. \ref{fig:pd}), i.e., we set the parameter $N= N_\delta$. The scaling dimensions of the stress tensor and symmetry current are
\begin{align}
    \Delta_T &=4-\epsilon+\tilde{\delta} \left(\frac{3 \epsilon ^2}{64}+\frac{\epsilon }{24}+\frac{1}{2}\right)- {\tilde{\delta}}^2 \left(\frac{1}{48}+\frac{23 \epsilon }{576}\right) + {\tilde{\delta}}^3 \frac{1}{96} ,\\
    \Delta_J &=3-\epsilon+\tilde{\delta} \left(\frac{29 \epsilon ^2}{576}+\frac{\epsilon }{24}+\frac{1}{2}\right)-{\tilde{\delta}}^2 \left(\frac{1}{48}+\frac{3 \epsilon }{64}\right)+ {\tilde{\delta}}^3\frac{1}{72}.
\end{align}
A key point in above results is that in the region $\delta\geqslant\delta^*$, $\Delta_{T/J}$ cross and go below the unitary bounds. 

Now consider more general case with $N$ goes away from the line $N=N_\delta+\delta n$. The scaling dimensions of the stress tensor and symmetry current in the chiral LRFP are
\begin{align}
    \Delta_T &=4-\epsilon+\tilde{\delta} \left(0.5+\left(0.0000745317 \text{$\delta $n}^{3/2}+0.000040306 \text{$\delta $n}^2-0.00136083 \text{$\delta $n}+0.0416667\right) \epsilon  \right.  \nn\\
    & \hspace{1.5cm} \left. +\left(0.000151951 \text{$\delta $n}^{3/2}+0.000199434 \text{$\delta $n}^2-0.00392372 \text{$\delta $n}+0.00130976 \sqrt{\text{$\delta $n}}+0.046875\right) \epsilon ^2 \right) \nn\\
    & \hspace{1.2cm} +{\tilde{\delta}}^2 \left(-0.0000372659 \text{$\delta $n}^{3/2}-0.000020153 \text{$\delta $n}^2+0.000680414 \text{$\delta $n}-0.0208333 \right.\nn\\
    & \hspace{1.5cm}\left. +\left(-0.000142635 \text{$\delta $n}^{3/2}-0.000196248 \text{$\delta $n}^2+0.00375362 \text{$\delta $n}-0.00130976 \sqrt{\text{$\delta $n}}-0.0399306\right) \epsilon\right) \nn\\
    &\hspace{1.2cm}+\tilde{\delta}^3\left(0.0000423692 \text{$\delta $n}^{3/2}+0.0000626169 \text{$\delta $n}^2-0.0011567 \text{$\delta $n}+0.000436587 \sqrt{\text{$\delta $n}}+0.0104167\right) , \label{eq:TdeltaN}\\
    \Delta_J &=3-\epsilon+\tilde{\delta} \left(0.5+\left(0.0000745317 \text{$\delta $n}^{3/2}+0.000040306 \text{$\delta $n}^2-0.00136083 \text{$\delta $n}+0.0416667\right) \epsilon  \right.  \nn\\
    & \hspace{1.5cm} \left. +\left(0.000158162 \text{$\delta $n}^{3/2}+0.000202793 \text{$\delta $n}^2-0.00403712 \text{$\delta $n}+0.00130976 \sqrt{\text{$\delta $n}}+0.0503472\right) \epsilon ^2\right) \nn\\
    & \hspace{1.2cm} +{\tilde{\delta}}^2 \left(-0.0000372659 \text{$\delta $n}^{3/2}-0.000020153 \text{$\delta $n}^2+0.000680414 \text{$\delta $n}-0.0208333 \right.\nn\\
    & \hspace{1.5cm}\left. +\left(-0.000155057 \text{$\delta $n}^{3/2}-0.000202966 \text{$\delta $n}^2+0.00398042 \text{$\delta $n}-0.00130976 \sqrt{\text{$\delta $n}}-0.046875\right) \epsilon\right) \nn\\
    &\hspace{1.2cm}+\tilde{\delta}^3\left(0.0000485802 \text{$\delta $n}^{3/2}+0.0000659758 \text{$\delta $n}^2-0.00127011 \text{$\delta $n}+0.000436587 \sqrt{\text{$\delta $n}}+0.0138889\right) . \label{eq:JdeltaN}
\end{align}
Scaling dimensions of the stress tensor and symmetry current in antichiral LRFP can be obtained by flipping the signs of the odd order terms of $\sqrt{\delta n}$ in above formulas. 

\end{document}